# Maximizing the magnitude of Strictly Standardized Mean Difference of a Linear Combination

*Estimation, Classification Cutoffs, Confidence Intervals, and Connections to LDA, Logistic Regression, and the ROC Curve*

Technical Note

**Xiaohua Douglas Zhang**

*Department of Biostatistics, College of Public Health, University of Kentucky, Lexington, KY, USA*

## Abstract

The strictly standardized mean difference (SSMD) is an interpretable, unit-free effect size originally developed for quality control and hit selection in high-throughput screening and subsequently extended to effect-size assessment and error-rate control. Existing SSMD theory considers one variable at a time. Here, we extend SSMD to a linear combination of multiple variables and determine the coefficients that maximize the resulting SSMD. We show that this optimization can be formulated as a generalized Rayleigh quotient with a closed-form solution and identify conditions under which the optimal direction coincides with Fisher's linear discriminant. We derive method-of-moments, bias-corrected, and maximum-likelihood estimators of the maximized SSMD and construct confidence intervals based on Hotelling's T-squared statistic and the noncentral F distribution under different correlation and variance structures. We also characterize classification cutoffs for the resulting linear score and establish its relationship with the area under the receiver operating characteristic curve (AUROC). Simulations show that SSMD maximization coincides with Fisher's linear discriminant and logistic regression under equal covariance but diverges under unequal covariance and class imbalance. In paired designs, SSMD maximization uniquely incorporates cross-covariance information. Linear support vector machines are also evaluated as a modern comparator. This framework provides an interpretable approach to constructing multivariable biomarkers and signatures for high-throughput screening, cytokine profiling, metabolomics, salivary diagnostics, and continuous-monitoring applications.

## Contents

# 1. Introduction

## 1.1 Background: the strictly standardized mean difference

The strictly standardized mean difference (SSMD) was introduced to quantify the magnitude of the difference between two groups in a way that is both interpretable and robust [1, 2]. For two groups with means $\mu_1$, $\mu_2$ and variances $\sigma_1^2$, $\sigma_2^2$, the SSMD is the mean difference divided by the standard deviation of that difference, i.e., $(\mu_1 - \mu_2)/\sqrt{\sigma_1^2 + \sigma_2^2 - 2\sigma_{12}}$. Unlike a t-statistic, it does not grow with sample size; unlike a raw mean difference, it is unit-free; and unlike the p-value, it speaks directly to effect magnitude [3]. SSMD has a clean probabilistic reading through the $d^+$-probability that a random draw from one group exceeds a random draw from the other, and threshold conventions such as |SSMD| ≥ 2 for a strong effect have become standard in screening practice [2, 4, 24].

Since its introduction, SSMD has been developed into a broad methodology: for hit selection and ranking [2, 4], an application now embedded in genome-scale RNAi and CRISPR screening pipelines and in the reproducibility assessment of microphysiological systems [27–31]; for controlling false discovery and false non-discovery rates [8]; for sample-size and error-rate determination [15]; as a contrast variable giving consistent interpretation to effect sizes across designs [5, 6]; as a replacement for the problematic z-factor in quality control [9]; and, most recently, for marker-gene selection in single-cell RNA-seq analysis [32]. A recent line of work links SSMD directly to machine-learning metrics such as AUROC, sensitivity, specificity, and accuracy [22, 23], and the methodology is consolidated in a monograph on optimal high-throughput screening [7].

## 1.2 Why a linear combination

In modern assays the analyst rarely has a single readout. A cytokine panel reports dozens of analytes [10]; a salivary diagnostic combines several biomarkers [11]; metabolomics produces thousands of features; continuous glucose and respiratory monitoring yield rich multivariate signals [12, 13]. In each case the goal is the same: combine the measured variables into a single, interpretable score that separates two conditions — disease versus control, responder versus non-responder, pre versus post — as strongly as possible.

This motivates the central question of the present note. For a predictor vector $\boldsymbol{X}$ we form a linear score $L = \boldsymbol{X}^T\boldsymbol{\beta}$ and ask which coefficient vector $\boldsymbol{\beta}$ maximizes the magnitude of SSMD of the two group scores. The answer extends SSMD from a per-variable statistic to a multivariate signature, and it brings SSMD into direct contact with the classical theory of linear classification.

## 1.3 Relationship to classical linear classifiers

The question has a classical echo. In 1936 Fisher asked which linear function of several measurements would maximize the ratio of the between-group mean difference to the within-group standard deviation, and answered it with the linear discriminant function [16]. SSMD maximization replaces the *within-group* standard deviation in Fisher's ratio with the standard deviation of the group *difference*, which retains the cross-covariance term and so coincides with Fisher's direction in some regimes but not others. Understanding when the two agree, and how SSMD magnitude maximization relates to logistic regression [20] and to margin-based classifiers such as the support vector machine [17], is one aim of this note. Because SSMD is tied to AUROC by AUROC = Φ(SSMD) [4, 22], the maximized score also inherits an interpretable ROC summary and a principled operating point [21].

### 1.4 Contributions and outline

This note makes four contributions. (i) We formulate SSMD maximization for a linear combination as a generalized Rayleigh quotient and derive the optimal direction $\boldsymbol{\beta}^* \propto \boldsymbol{\Sigma}_*^{-1}(\boldsymbol{\mu}_1 - \boldsymbol{\mu}_2)$, identifying the cases in which it reduces to Fisher's discriminant [16, 19] (Sections 2–3). (ii) We give point estimators of the maximized SSMD by the method of moments, the UMVUE with the Hedges bias correction [18], and maximum likelihood, with exact and approximate confidence intervals from Hotelling's $T^2$ and the noncentral F-distribution (Section 4). (iii) We characterize the classification cutoff on the score (Section 5) and compare SSMD maximization with Fisher LDA, logistic regression, and a linear SVM, including the regimes in which they diverge (Section 6). (iv) We connect the maximized SSMD to AUROC and standard operating points and illustrate the framework by simulation [22]. The remainder of the note develops the setup and notation (Section 2), the optimal direction (Section 3), estimation and inference (Section 4), classification cutoffs (Section 5), the comparison with Fisher LDA, logistic regression, and linear SVM—including the regimes in which they diverge (Section 6)—and closes with a discussion (Section 7).

## 2. Problem Setup and Notation

We observe data on *n* subjects, each with *k* predictors, belonging to one of two groups, $G_1$ or $G_2$. For subject *i* the predictor vector is

$$\boldsymbol{X}_i = (X_{i1}, X_{i2}, \cdots, X_{ik})^T, i = 1,2,\cdots,n$$

We form a linear score

$$\boldsymbol{L} = \beta_0 + \sum_{j=1}^{k} \beta_j \boldsymbol{X}_j = \beta_0 + \boldsymbol{X}^T\boldsymbol{\beta}, \qquad \boldsymbol{\beta} = (\beta_1, \cdots, \beta_k)^T$$

Let $L_1$ and $L_2$ be the score in each group and $D = L_1 - L_2$ their difference.

Notation:

- $n_1$, $n_2$: group sizes ($n = n_1 + n_2$).
- $\boldsymbol{\mu}_1$, $\boldsymbol{\mu}_2$: population predictor means; estimates $\bar{\boldsymbol{X}}_1$, $\bar{\boldsymbol{X}}_2$.
- $\boldsymbol{\Sigma}_1$, $\boldsymbol{\Sigma}_2$: within-group covariance matrices; estimates $\mathbf{S}_1$, $\mathbf{S}_2$.
- $\boldsymbol{\Sigma}_{12}$: cross-covariance between the group scores (nonzero only for paired groups); estimate $\mathbf{S}_{12}$.
- $\mu_{L1}, \mu_{L2}$: population means of $L_1$ and $L_2$.
- $\sigma_{L1}^2, \sigma_{L2}^2$: population variance of $L_1$ and $L_2$.
- $\sigma_{L12}$: population covariance between $L_1$ and $L_2$.

**Objective:** choose $\boldsymbol{\beta}$ to maximize the SSMD of the two group scores,

$$\mathrm{SSMD}_D = \frac{\mu_d}{\sigma_D} = \frac{\mu_{L1} - \mu_{L2}}{\sqrt{\sigma_{L1}^2 + \sigma_{L2}^2 - 2\sigma_{L12}}}$$

The denominator is the standard deviation of the **difference**, so it keeps the covariance term $-2\sigma_{L12}$ [3].

The population SSMD takes one of three forms, depending on the variance and correlation structure of the two group scores. Situation (a) is the independent, equal-variance case ($\sigma_{L1}^2 =$ $\sigma_{L2}^2 = \sigma_L^2, \sigma_{L12} = 0$). Situation (b) assumes independence but allows unequal variances ($\sigma_{L1}^2 \neq \sigma_{L2}^2, \sigma_{L12} = 0$). Situation (c) is the general case, in which the covariance between the two group scores is nonzero ($\sigma_{L12} \neq 0$), as occurs, for example, in paired-study designs. The corresponding population SSMDs are,

$$\mathrm{SSMD}_{(a)} = \frac{\mu_{L1} - \mu_{L2}}{\sqrt{2}\,\sigma_L}$$

$$\mathrm{SSMD}_{(b)} = \frac{\mu_{L1} - \mu_{L2}}{\sqrt{\sigma_{L1}^2 + \sigma_{L2}^2}}$$

$$\mathrm{SSMD}_{(c)} = \frac{\mu_{L1} - \mu_{L2}}{\sqrt{\sigma_{L1}^2 + \sigma_{L2}^2 - 2\sigma_{L12}}}$$

# 3. Estimating the Coefficients of a Linear Combination by Maximizing SSMD

## 3.1 The SSMD as a generalized Rayleigh quotient

Because $\beta_0$ is shared, it cancels in the mean difference:

$$\mu_d = \boldsymbol{\beta}^T(\boldsymbol{\mu}_1 - \boldsymbol{\mu}_2)$$

Propagating the linear map through the joint covariance of the two group scores gives the variance of *D*:

$$\sigma_d^2 = \boldsymbol{\beta}^T(\boldsymbol{\Sigma}_1 + \boldsymbol{\Sigma}_2 - \boldsymbol{\Sigma}_{12} - \boldsymbol{\Sigma}_{12}^T)\boldsymbol{\beta}$$

Writing $\boldsymbol{m} = \boldsymbol{\mu}_1 - \boldsymbol{\mu}_2$ and $\boldsymbol{\Sigma}_* = \boldsymbol{\Sigma}_1 + \boldsymbol{\Sigma}_2 - \boldsymbol{\Sigma}_{12} - \boldsymbol{\Sigma}_{12}^T$, the objective is a **generalized Rayleigh quotient**:

$$\mathrm{SSMD}(\boldsymbol{\beta}) = \frac{\boldsymbol{\beta}^{\mathrm{T}}\boldsymbol{m}}{\sqrt{\boldsymbol{\beta}^T\boldsymbol{\Sigma}_*\boldsymbol{\beta}}}$$

## 3.2 The maximizer

Maximizing the generalized Rayleigh quotient $\boldsymbol{\beta}^{\mathrm{T}}\boldsymbol{m}/\sqrt{\boldsymbol{\beta}^{\mathrm{T}}\boldsymbol{\Sigma}_*\boldsymbol{\beta}}$ over non-zero $\boldsymbol{\beta}$ with non-zero $\boldsymbol{m}$ and symmetric positive-definite $\boldsymbol{\Sigma}_*$ yields, up to positive scale,

$$\boldsymbol{\beta}^* \propto \boldsymbol{\Sigma}_*^{-1}\boldsymbol{m} = (\boldsymbol{\Sigma}_1 + \boldsymbol{\Sigma}_2 - \boldsymbol{\Sigma}_{12} - \boldsymbol{\Sigma}_{12}^T)^{-1}\,(\boldsymbol{\mu}_1 - \boldsymbol{\mu}_2)$$

Equivalently, the square of this ratio, $\boldsymbol{\beta}^{\mathrm{T}}\boldsymbol{m}\boldsymbol{m}^{\boldsymbol{T}}\boldsymbol{\beta}/\boldsymbol{\beta}^{\mathrm{T}}\boldsymbol{\Sigma}_*\boldsymbol{\beta}$, is a generalized Rayleigh quotient with rank-one numerator matrix $\boldsymbol{m}\boldsymbol{m}^{\boldsymbol{T}}$. Maximizing the square determines $\boldsymbol{\beta}^*$ up to any nonzero scale, and the orientation is fixed by maximizing the ratio itself.

The solution follows solely from the linearity of expectation and second-moment propagation. It requires only finite first and second moments and a positive-definite $\boldsymbol{\Sigma}_*$, with no assumption that the components of the predictor vector $\boldsymbol{X}$ are uncorrelated, that the two groups are independent, or that $\boldsymbol{X}$ is normally distributed.

Set $\boldsymbol{\beta}^* = a\,\boldsymbol{\Sigma}_*^{-1}\boldsymbol{m} = a\,(\boldsymbol{\Sigma}_1 + \boldsymbol{\Sigma}_2 - \boldsymbol{\Sigma}_{12} - \boldsymbol{\Sigma}_{12}^T)^{-1}\,(\boldsymbol{\mu}_1 - \boldsymbol{\mu}_2)$ for a nonzero real constant a. Then

$$\mathrm{SSMD}(\boldsymbol{\beta}^*) = \frac{\boldsymbol{\beta}^{*\mathrm{T}}(\boldsymbol{\mu}_1 - \boldsymbol{\mu}_2)}{\sqrt{\boldsymbol{\beta}^{*T}\boldsymbol{\Sigma}_*\boldsymbol{\beta}^*}} = \frac{(a\,\boldsymbol{\Sigma}_*^{-1}\boldsymbol{m})^T\boldsymbol{m}}{\sqrt{(a\,\boldsymbol{\Sigma}_*^{-1}\boldsymbol{m})^T\boldsymbol{\Sigma}_* a\boldsymbol{\Sigma}_*^{-1}\boldsymbol{m}}} = \mathrm{sign}(a)\cdot\sqrt{\boldsymbol{m}^T\boldsymbol{\Sigma}_*^{-1}\boldsymbol{m}}$$

Consequently, over all $\boldsymbol{\beta} \neq \boldsymbol{0}$, the SSMD of a linear combination satisfies

$$-\sqrt{\boldsymbol{m}^T\boldsymbol{\Sigma}_*^{-1}\boldsymbol{m}} \leq \mathrm{SSMD}(\boldsymbol{\beta}) \leq \sqrt{\boldsymbol{m}^T\boldsymbol{\Sigma}_*^{-1}\boldsymbol{m}}$$

The upper bound is attained when $a > 0$ and the lower bound when $a < 0$. The magnitude of $a$ is immaterial — only its sign must be fixed — and this choice orients the combined score rather than determining which extremum is attained: $a > 0$ gives the maximizer, $a < 0$ the minimizer.

Recall the definition of the SSMD of a linear combination evaluated at $\boldsymbol{\beta}^*$,

$$\mathrm{SSMD_D}(\boldsymbol{\beta}^*) = \frac{\mu_{L1(\boldsymbol{\beta}^*)} - \mu_{L2(\boldsymbol{\beta}^*)}}{\sqrt{\sigma_{L1(\boldsymbol{\beta}^*)}^2 + \sigma_{L2(\boldsymbol{\beta}^*)}^2 - 2\sigma_{L12(\boldsymbol{\beta}^*)}}}$$

Because $\mu_{L1(\boldsymbol{\beta}^*)} - \mu_{L2(\boldsymbol{\beta}^*)} = (a\,\boldsymbol{\Sigma}_*^{-1}\boldsymbol{m})^T\boldsymbol{m}$, the default rule — choose the sign of $a$ so that the numerator of the expression above is positive — is exactly the choice $a > 0$, which attains the maximum. The sign may also be chosen to suit the specific aims of a study. For example, in a high-content screen one may choose the sign so that the mean of the linear combination is larger in an up-regulated positive control than in the negative control, or smaller in a down-regulated positive control than in the negative control.

Replacing population quantities by sample estimates gives the estimator

$$\widehat{\boldsymbol{\beta}} \propto (\boldsymbol{S}_1 + \boldsymbol{S}_2 - \boldsymbol{S}_{12} - \boldsymbol{S}_{12}^T)^{-1}(\overline{\boldsymbol{X}}_1 - \overline{\boldsymbol{X}}_2)$$

The SSMD is scale-invariant in $\boldsymbol{\beta}$, so only the direction is identified; $\widehat{\boldsymbol{\beta}}$ may be normalized to unit length. The corresponding estimated maximal SSMD of the linear combination is,

$$\widehat{\mathrm{SSMD}}(\boldsymbol{\beta}^*) = \mathrm{sign}(a)\sqrt{(\overline{\boldsymbol{X}}_1 - \overline{\boldsymbol{X}}_2)^T(\boldsymbol{S}_1 + \boldsymbol{S}_2 - \boldsymbol{S}_{12} - \boldsymbol{S}_{12}^T)^{-1}(\overline{\boldsymbol{X}}_1 - \overline{\boldsymbol{X}}_2)}$$

**Special case 1 (situation b: independent groups, unequal covariances).** If $\boldsymbol{\Sigma}_{12} = \boldsymbol{0}$, then $\boldsymbol{\Sigma}_* = \boldsymbol{\Sigma}_1 + \boldsymbol{\Sigma}_2$ and

$$\boldsymbol{\beta}^* \propto (\boldsymbol{\Sigma}_1 + \boldsymbol{\Sigma}_2)^{-1}\,(\boldsymbol{\mu}_1 - \boldsymbol{\mu}_2)$$

which is the optimal linear discriminant direction under unequal covariances (the heteroscedastic / Anderson–Bahadur linear direction) [19]. Note this is not Fisher's quadratic discriminant: when $\Sigma_1 \neq \Sigma_2$ the Bayes-optimal rule (QDA) has a genuinely quadratic boundary and no single direction vector; the linear score above is optimal only within the class of linear projections (cf. Section6.3).

Replacing population quantities by sample estimates gives the estimator

$$\widehat{\boldsymbol{\beta}} \propto (\boldsymbol{S}_1 + \boldsymbol{S}_2)^{-1}(\overline{\boldsymbol{X}}_1 - \overline{\boldsymbol{X}}_2)$$

**Special case 2 (situation a: independent groups, equal covariance).** If $\boldsymbol{\Sigma}_{12} = \boldsymbol{0}$ and $\boldsymbol{\Sigma}_1 = \boldsymbol{\Sigma}_2 = \boldsymbol{\Sigma}$, then $\boldsymbol{\Sigma}_* = 2\boldsymbol{\Sigma}$ and

$$\boldsymbol{\beta}^* \propto \boldsymbol{\Sigma}^{-1}\,(\boldsymbol{\mu}_1 - \boldsymbol{\mu}_2)$$

Replacing population quantities by sample estimates gives the estimator

$$\widehat{\boldsymbol{\beta}} \propto \boldsymbol{S}_p^{-1}(\overline{\boldsymbol{X}}_1 - \overline{\boldsymbol{X}}_2)$$

which is exactly Fisher's discriminant direction (Section 6). Here $\mathbf{S}_p$ is the pooled within-group covariance matrix, formed by combining the two within-group estimates with their degrees of freedom:

$$\boldsymbol{S}_p = \frac{(n_1 - 1)\boldsymbol{S}_1 + (n_2 - 1)\boldsymbol{S}_2}{n_1 + n_2 - 2}$$

This is the same pooled scatter that Fisher's LDA uses.

### 3.3 The intercept $\beta_0$

The intercept cancels in $L_1 - L_2$, so it does not affect the SSMD and is a free centering choice. A convenient convention sets the grand mean of the score to zero,

$$\beta_0 = \overline{\boldsymbol{L}} - \widehat{\boldsymbol{\beta}}^T\overline{\boldsymbol{X}}, \qquad \overline{\boldsymbol{X}} = \frac{n_1\overline{\boldsymbol{X}}_1 + n_2\overline{\boldsymbol{X}}_2}{n_1 + n_2}$$

which gives $\beta_0 = 0$ when the score is centered at the overall mean. For classification one instead places $\beta_0$ so the threshold sits between the group score means (Section 5).

## 4. Estimate and Confidence Interval of the Maximal SSMD of a Linear Combination

### 4.1 Optimized-$\boldsymbol{\beta}^*$ case

Section 3 showed that the SSMD of the score $\boldsymbol{L} = \boldsymbol{X}^T\boldsymbol{\beta}$ is maximized at $\boldsymbol{\beta}^* = a\,\boldsymbol{\Sigma}_*^{-1}(\boldsymbol{\mu}_1 - \boldsymbol{\mu}_2)$, and that its maximal value is $\mathrm{SSMD}(\boldsymbol{\beta}^*) = \mathrm{sign}(a) \cdot \sqrt{(\boldsymbol{\mu}_1 - \boldsymbol{\mu}_2)^T\boldsymbol{\Sigma}_*^{-1}(\boldsymbol{\mu}_1 - \boldsymbol{\mu}_2)}$. For a *fixed* direction $\beta$ the score is univariate and $\mathrm{SSMD}(\boldsymbol{\beta}) = \mathrm{sign}(a) \cdot \frac{\boldsymbol{\beta}^{\mathrm{T}}\boldsymbol{m}}{\sqrt{\boldsymbol{\beta}^T\boldsymbol{\Sigma}_*\boldsymbol{\beta}}}$ is a ratio of a normal mean difference to an independent chi-type denominator, so its exact sampling law is a *noncentral t* — the single-direction construction of Zhang [7]. The maximal SSMD is not such a ratio: it is the square root of a *quadratic form* in the estimated mean difference, standardized by an estimated covariance. Its natural sample analogue is therefore Hotelling's $T^2$, and its exact sampling law is a *noncentral F* [19]. Using the noncentral *F* rather than the noncentral *t* is precisely what accounts for $\boldsymbol{\beta}^*$ having been estimated from the same data — the optimism that an interval built for a pre-specified direction ignores. Throughout we write $\widehat{\boldsymbol{m}} = \overline{\boldsymbol{X}}_1 - \overline{\boldsymbol{X}}_2$ and let $\widehat{\boldsymbol{\Sigma}}_*$ denote the estimate of $\boldsymbol{\Sigma}_*$ appropriate to the design (pooled, unpooled, or paired).

### 4.1.1 Point estimation: plug-in, bias correction, and MLE

**Plug-in (method of moments).** Replacing $\boldsymbol{m}$ and $\boldsymbol{\Sigma}_*$ by sample quantities gives

$$\widehat{\text{SSMD}}(\boldsymbol{\beta}^*) = \text{sign}(a) \cdot \sqrt{(\overline{\boldsymbol{X}}_1 - \overline{\boldsymbol{X}}_2)^T \ \widehat{\boldsymbol{\Sigma}}_*{}^{-1} (\overline{\boldsymbol{X}}_1 - \overline{\boldsymbol{X}}_2)}$$

, which equals $\sqrt{(\overline{\boldsymbol{X}}_1 - \overline{\boldsymbol{X}}_2)^T \ \mathbf{S}_\text{p}{}^{-1} (\overline{\boldsymbol{X}}_1 - \overline{\boldsymbol{X}}_2)/2} = \widehat{\text{D}}/\sqrt{2}$ and $S_\text{p}$ is the pooled within-group sample covariance in situation (a), $\sqrt{(\overline{\boldsymbol{X}}_1 - \overline{\boldsymbol{X}}_2)^T \ (\mathbf{S}_1 + \mathbf{S}_2)^{-1} (\overline{\boldsymbol{X}}_1 - \overline{\boldsymbol{X}}_2)}$ in situation (b) and $\sqrt{\overline{\boldsymbol{d}}^T \, \mathbf{S}_\text{d}^{-1} \, \overline{\boldsymbol{d}}}$ in situation (c).

**Finite-sample bias.** The plug-in over-estimates the maximal SSMD, and — unlike the single-direction case — the bias grows with the dimension *k*. Under within-group normality, $\widehat{D}^2 = \left(\overline{\boldsymbol{X}}_1 - \overline{\boldsymbol{X}}_2\right)^T \ \mathbf{S}_\text{p}{}^{-1} (\overline{\boldsymbol{X}}_1 - \overline{\boldsymbol{X}}_2)$ (situation a) satisfies

$$\text{E}\left|\widehat{D}^2\right| = \frac{\nu}{\nu - k - 1}\left(D^2 + \frac{\text{k}(\text{n}_1 + \text{n}_2)}{n_1 n_2}\right), \qquad \nu = n_1 + n_2 - 2$$

so an exactly unbiased estimator of the squared population distance is

$$\widehat{\text{D}}_u^2 = \frac{\nu}{\nu - k - 1}\widehat{D}^2 - \frac{\text{k}(\text{n}_1 + \text{n}_2)}{n_1 n_2}$$

with the bias-corrected maximal SSMD: $\sqrt{max(\widehat{D}_u^2, 0)/2}$ in situation (a). The truncation at zero is required because $\widehat{D}_u^2$ can be negative when the true separation is small; because the square root is nonlinear, no estimator of the maximal SSMD is exactly unbiased even when $\widehat{D}_u^2$ is. Both the multiplicative shrinkage $(\nu - k - 1)/\nu$ and the additive term $k(n_1+n_2)/(n_1 n_2)$ vanish as $\nu \to \infty$, so all estimators are consistent and share the same probability limit.

**Maximum likelihood.** Under within-group normality, the MLE divides each scatter matrix by $n_\text{k}$ rather than $n_\text{k} - 1$, which inflates $\widehat{D}^2$ further; as in the single-direction case [7, 15], the finite-sample bias ordering is MLE > plug-in > bias-corrected.

### 4.1.2 Exact and approximate confidence intervals via the noncentral F-distribution

An exact interval follows by inverting Hotelling's $T^2$, whose scaled form has a noncentral *F* distribution [19]. We give the two exact designs first, then the approximate one.

**Situation (a): equal covariance, independent groups (two-sample $T^2$).** With $S_\text{p}$ the pooled covariance, form

$$T^2 = \frac{n_1 n_2}{n_1 + n_2}\widehat{\boldsymbol{m}}^T \boldsymbol{S}_p{}^{-\mathbf{1}} \widehat{\boldsymbol{m}}, \qquad F = \frac{\nu - k + 1}{k\nu} T^2 \sim F_{k, \nu - k + 1}(\lambda)$$

an *F* on *k* and $\nu - k + 1$ degrees of freedom with noncentrality

$$\lambda = \frac{n_1 n_2}{n_1 + n_2} D^2 = \frac{2n_1 n_2}{n_1 + n_2} \mathrm{SSMD}(\boldsymbol{\beta}^*)^2$$

Inverting the noncentral-*F* CDF at the observed *F* gives an exact 1 − *α* confidence set $[\lambda_{\mathrm{lo}}, \lambda_{\mathrm{hi}}]$ for the noncentrality,

$$\lambda_{\mathrm{lo}}: \mathrm{P}(F \le F_{\mathrm{obs}} | \lambda_{\mathrm{lo}}) = 1 - \alpha/2, \qquad \lambda_{\mathrm{hi}}: \mathrm{P}(F \le F_{\mathrm{obs}} | \lambda_{\mathrm{hi}}) = \alpha/2$$

with$\lambda_{\mathrm{lo}}$ set to 0 whenever the observed *F* is too small to exclude it. Mapping the endpoints back through the noncentrality relation yields the interval for the maximal SSMD,

$$\left( \sqrt{\frac{n_1 + n_2}{2n_1 n_2} \lambda_{\mathrm{lo}}}, \ \sqrt{\frac{n_1 + n_2}{2n_1 n_2} \lambda_{\mathrm{hi}}} \right)$$

**Situation (c): paired or correlated groups (one-sample $T^2$).** Form the subject-level difference vectors $\boldsymbol{d_i} = \boldsymbol{X_{i1}} - \boldsymbol{X_{i2}}$, with sample mean $\overline{\boldsymbol{d}}$ and covariance $\mathbf{S}_{\mathrm{d}}$. Since $cov(\boldsymbol{d_i}) = \boldsymbol{\Sigma}_*$, Hotelling's $T^2$ is $T^2 = n\overline{\boldsymbol{d}}^T \mathbf{S}_{\mathrm{d}} \overline{\boldsymbol{d}}$, and

$$F = \frac{n-k}{k(n-1)} T^2 \sim F_{k,n-k}(\lambda) \ \text{ where } \lambda = n\, \mathrm{SSMD}\left(\beta_*\right)^2$$

The interval is obtained by the same inversion, with the simpler map $\mathrm{SSMD}_{lo,hi} = \sqrt{\lambda_{lo,hi}/n}$. This construction uses the cross-covariance exactly and is the design in which the SSMD linear combination has no fixed-direction counterpart.

**Situation (b): unequal covariance, independent groups (multivariate Behrens–Fisher).** Here $\boldsymbol{\Sigma}_* = \boldsymbol{\Sigma_1} + \boldsymbol{\Sigma_2}$ admits no exact Hotelling law. Using the unpooled second-moment matrix, the generalized statistic

$$T_*^2 = \widehat{\boldsymbol{m}}^T \left( \frac{\boldsymbol{S}_1}{n_1} + \frac{\boldsymbol{S}_2}{n_2} \right)^{\boldsymbol{-1}} \widehat{\boldsymbol{m}}, \qquad F = \frac{d-k+1}{kd} T_*^2 \sim F_{k,d-k+1}(\lambda)$$

is referred to an *F* with Nel–van der Merwe effective degrees of freedom *d* [25], and the interval is formed by the same inversion. When $n_1 = n_2 = n$ the noncentrality reduces exactly to $\lambda = n\, \mathrm{SSMD}(\boldsymbol{\beta}^*)^2$; for unequal sample sizes, the map $\lambda = \boldsymbol{m}^T \left( \frac{\boldsymbol{\Sigma}_1}{n_1} + \frac{\boldsymbol{\Sigma}_2}{n_2} \right)^{-1} \boldsymbol{m}$ is only approximately proportional to $\mathrm{SSMD}(\boldsymbol{\beta}^*)^2$, and coverage is approximate. Because the plug-in is upward biased here, we recommend the bias-corrected bootstrap of §4.4 as the primary tool for situation (b).

**Reduction to one dimension.** When *k* = 1 the noncentral $F_{1,\nu}(\lambda)$ is the square of a noncentral *t* with $\delta = \sqrt{\lambda}$, and the entire construction collapses to the exact single-direction noncentral-*t* interval of Zhang [7]: the fixed-*β* result is the one-dimensional special case of the maximal-SSMD interval. In simulation the noncentral-*F* intervals attain near-nominal coverage in situations (a) and (c) and slight under-coverage in situation (b).

### 4.1.3 Delta-method interval (large-sample)

For non-normal scores, or as a cross-check, a large-sample interval follows from the delta method applied to the quadratic form. Treating $\boldsymbol{\Sigma}_*$ as fixed to leading order, with $\mathbf{A} = \boldsymbol{\Sigma}_*^{-1}$ and gradient $\frac{\partial(\widehat{\boldsymbol{m}}^T \boldsymbol{A}\, \widehat{\boldsymbol{m}})}{\partial \boldsymbol{m}} = 2\boldsymbol{Am},$

$$\mathrm{Var}\left( \widehat{\mathrm{SSMD}}(\boldsymbol{\beta}^*) \right) \approx \frac{\widehat{\boldsymbol{m}}^T \boldsymbol{A}\, \widehat{\boldsymbol{V}}\, \boldsymbol{A}\, \widehat{\boldsymbol{m}}}{\widehat{\boldsymbol{m}}^T \boldsymbol{A}\, \widehat{\boldsymbol{m}}}$$

with $\widehat{\boldsymbol{V}} = \frac{\boldsymbol{S_1}}{\boldsymbol{n_1}} + \frac{\boldsymbol{S_2}}{\boldsymbol{n_2}}$ for independent groups or $\widehat{\boldsymbol{V}} = \frac{\boldsymbol{S_D}}{\boldsymbol{n}}$ for paired groups, giving the Wald interval $\widehat{\mathrm{SSMD}}(\boldsymbol{\beta}^*) \pm z_{1-\frac{\alpha}{2}} \cdot \sqrt{\mathrm{Var}\left(\widehat{\mathrm{SSMD}}(\boldsymbol{\beta}^*)\right)}$. This term captures only the variability of the mean difference; the noncentral-*F* intervals additionally account for estimating $\boldsymbol{\Sigma}_*$, so the delta interval is somewhat anticonservative in small samples.

#### 4.1.4 Bias-corrected bootstrap

Because $\boldsymbol{\beta}^*$ and $\boldsymbol{\Sigma}_*$ are estimated jointly, the most transparent small-sample interval resamples subjects with replacement within each group (holding $n_1$, $n_2$ fixed) and, on each resample, recomputes the full pipeline — $\boldsymbol{\beta}^*$, the projected score, and the plug-in maximal SSMD $\sqrt{\widehat{\boldsymbol{m}}^T \boldsymbol{\Sigma}_*^{-1}\, \widehat{\boldsymbol{m}}}$. Because the plug-in carries the upward Mahalanobis bias of §4.1, the plain percentile interval inherits that bias and under-covers; a bias-corrected-and-accelerated (BCa) interval is preferred. The bootstrap is the recommended default in situation (b), where no exact interval exists, and a useful check on the noncentral-*F* intervals in situations (a) and (c).

### 4.2 Fixed-$\boldsymbol{\beta}$ case

This section gives point estimators and confidence intervals for the SSMD of a fixed linear combination $L = \boldsymbol{X}^T\boldsymbol{\beta}$ with direction $\boldsymbol{\beta}$ held fixed (the direction is identified only up to scale; see Section 3.2). Writing the per-group score mean and variance as $\mu_{Lk} = \boldsymbol{\beta^T \mu_k}$ and $\sigma_{Lk}^2 = \boldsymbol{\beta^T \Sigma_k \beta}$, and the cross term as $\sigma_{L12} = \boldsymbol{\beta^T \Sigma_{12} \beta}$, we estimate $\boldsymbol{\beta}$ from the projected scores, with sample quantities

$$L = \boldsymbol{X}^T\boldsymbol{\beta}, \mu_{Lk} = \boldsymbol{\beta^T \mu_k},\ \sigma_{Lk}^2 = \boldsymbol{\beta^T \Sigma_k \beta},\ \sigma_{L12} = \boldsymbol{\beta^T \Sigma_{12} \beta}$$

where $\bar{L}, s_{Lk}^2$ are the sample mean and variance of the $n_k$ projected scores in group *k*, and $S_{L12}$ is the sample cross-covariance of the paired scores (set $S_{L12} = 0$ for independent groups).

#### 4.2.1 Point estimation: method of moments, UMVUE, and MLE

**Method of moments (MM).** The MM estimator substitutes unbiased sample moments into the population definition. In the general case (situation c described in Section 2),

$$\widehat{\mathrm{SSMD}}_{\mathrm{MM}} = \frac{\bar{L}_1 - \bar{L}_2}{\sqrt{s_{L1}^2 + s_{L2}^2 - 2s_{L12}}}$$

which specializes, for the independent-group situations (a) and (b), to

$$\widehat{\mathrm{SSMD}}_{\mathrm{MM}}^{(\mathrm{a})} = \frac{\bar{L}_1 - \bar{L}_2}{\sqrt{2}\, s_p}, \qquad \text{where the pooled variance } s_p^2 = \frac{(n_1 - 1)s_{L1}^2 + (n_2 - 1)s_{L2}^2}{n_1 + n_2 - 2}$$

and

$$\widehat{\mathrm{SSMD}}_{\mathrm{MM}}^{(\mathrm{b})} = \frac{\bar{L}_1 - \bar{L}_2}{\sqrt{s_{L1}^2 + s_{L2}^2}}.$$

**Maximum likelihood (MLE).** Under within-group normality the MLE replaces each unbiased variance $s_{Lk}^2$ by its maximum-likelihood counterpart $\tilde{\sigma}_{Lk}^2 = (n_k - 1)s_{Lk}^2 / n_k$ (and analogously for the cross term $\tilde{\sigma}_{L12}$), giving

$$\widehat{\mathrm{SSMD}}_{\mathrm{MLE}} = \frac{\bar{L}_1 - \bar{L}_2}{\sqrt{\tilde{\sigma}_{L1}^2 + \tilde{\sigma}_{L2}^2 - 2\,\tilde{\sigma}_{L12}}}.$$

Because the ML variances are smaller than the unbiased ones, the MLE has the largest positive bias of the three estimators in finite samples.

**Uniformly minimum-variance unbiased estimator (UMVUE).** The MM and ML estimators both over-estimate SSMD in small samples because $\mathrm{E}\left[\frac{1}{s}\right] \neq \frac{1}{\mathrm{E}[s]}$. Under normality, this bias is removed exactly by the multiplicative constant

$$K_\nu = \sqrt{\frac{2}{\nu}}\frac{\Gamma(\frac{\nu}{2})}{\Gamma(\frac{\nu-1}{2})}$$

the same correction factor used for the UMVUE of a single-direction SSMD [7] and for Hedges' *g* [18]. In situation (a) the unbiased estimator is

$$\widehat{\mathrm{SSMD}}_{\mathrm{UMVUE}}^{(\mathrm{a})} = K_\nu \widehat{\mathrm{SSMD}}_{\mathrm{MM}}^{(\mathrm{a})} = K_\nu \frac{\bar{L}_1 - \bar{L}_2}{\sqrt{2}\, s_p}, \qquad \text{where } \nu = n_1 + n_2 - 2$$

In situations (b) and (c) the variances are not pooled, so the exact degrees of freedom are replaced by the Welch–Satterthwaite effective value $\nu_{\mathrm{eff}}$ where

$$\nu_{\mathrm{eff}} = \frac{(s_{L1}^2 + s_{L2}^2)^2}{\frac{s_{L1}^4}{n_1 - 1} + \frac{s_{L2}^4}{n_2 - 1}}$$

that matches the variance of the denominator:

$$\widehat{\mathrm{SSMD}}_{\mathrm{UMVUE}}^{(\mathrm{b})} = K_{\nu_{\mathrm{eff}}} \frac{\bar{L}_1 - \bar{L}_2}{\sqrt{s_{L1}^2 + s_{L2}^2}}$$

$$\widehat{\mathrm{SSMD}}_{\mathrm{UMVUE}}^{(\mathrm{c})} = K_{\nu_{\mathrm{eff}}} \frac{\bar{L}_1 - \bar{L}_2}{\sqrt{s_{L1}^2 + s_{L2}^2 - 2 s_{L12}}}$$

All three estimators are consistent and share the same probability limit β; they differ only by an *O*(1/ν) scalar. The finite-sample bias ordering is MLE > MM > UMVUE, with the UMVUE unbiased to machine precision as shown by Zhang [3,6,7] for a contrast variable, and all three converging as $\nu \to \infty$.

### 4.2.2 Exact and approximate confidence intervals via the noncentral t-distribution

Because each estimator is a (scaled) ratio of a normal mean difference to an independent χ-type denominator, an exact small-sample interval follows from the noncentral *t*-distribution rather than from a normal approximation. Following the logic described in Chapter 7 of Zhang [7], we can give the derivation for situation (b) — zero correlation, unequal variance — and then state the parallel results for (a) and (c).

### 4.2.3 Delta-method interval (large-sample)

Using the asymptotical property of maximal likelihood estimate similarly as shown in Zhang [1], treating numerator and denominator as approximately independent for large $n_1$, $n_2$,

$$\mathrm{Var}\left(\widehat{\mathrm{SSMD}}_D\right) \approx \frac{\mathrm{Var}(\hat{\mu}_d)}{\hat{\sigma}_d^2} + \frac{\hat{\mu}_d^2\,\mathrm{Var}\left(\hat{\sigma}_d^2\right)}{4\hat{\sigma}_d^6}$$

with the *1−α* interval

$$\widehat{\mathrm{SSMD}}_D \pm z_{1-\frac{\alpha}{2}}\sqrt{\mathrm{Var}\left(\widehat{\mathrm{SSMD}}_D\right)}$$

#### 4.2.4 Classical closed form (Cohen's d)

**Caveat.** The familiar closed form Caveat.

$$\mathrm{SE}\left(\hat{d}\right) = \sqrt{\frac{1}{n_1} + \frac{1}{n_2} + \frac{d^2}{2(n_1 + n_2)}}$$

is the large-sample SE of Cohen's d, whose denominator is a pooled or average SD rather than $\sqrt{\sigma_1 + \sigma_2}$. Under equal variances the two indices satisfy $\mathrm{SSMD} = \frac{d}{\sqrt{2}}$, so the corresponding SSMD expression halves the first term:

$$\mathrm{SE}\left(\widehat{\mathrm{SSMD}}\right) = \sqrt{\frac{1}{2}\left(\frac{1}{n_1} + \frac{1}{n_2}\right) + \frac{\mathrm{SSMD}^2}{2(n_1 + n_2)}}$$

Applying the Cohen's d form directly to SSMD overstates the standard error. Under unequal variances no fixed ratio holds, and the delta method or bootstrap should be used.

#### 4.2.5 Bootstrap (recommended for small samples)

1. Resample subjects with replacement within each group, keeping $n_1$, $n_2$ fixed.
2. Recompute $\hat{\boldsymbol{\beta}}$ and $\widehat{SSMD}_D$ on each resample.
3. Report the empirical *1−α* percentile interval; this also captures the optimism from estimating $\boldsymbol{\beta}$ on the same data.

## 5. Discriminating Decision Boundaries

Once $\boldsymbol{\beta}$ is fixed, classification reduces to thresholding the scalar score $L = \boldsymbol{X}^{\boldsymbol{T}}\boldsymbol{\beta}$. Within each group the score is normal:

$$L|G_1 \sim N(\mu_{L1}, \sigma_1^2), \qquad L|G_2 \sim N(\mu_{L2}, \sigma_2^2)$$

with

$$\mu_{Lk} = \boldsymbol{\beta}^{\boldsymbol{T}}\boldsymbol{\mu}_{\boldsymbol{k}}, \qquad \sigma_{Lk}^2 = \boldsymbol{\beta}^{\boldsymbol{T}}\boldsymbol{\Sigma}_{\boldsymbol{k}}\boldsymbol{\beta},$$

Take $\mu_{L1} > \mu_{L2}$ and classify into $G_1$ when $L > c$.

### 5.1 The Bayes-optimal cutoff for the SSMD direction

The SSMD maximization delivers the projection direction $\hat{\boldsymbol{\beta}}$, but it says nothing about where to place the threshold c. The threshold can be obtained by a separate, principled step: once the data are projected onto the SSMD axis, choose the cutoff that minimizes the expected misclassification (Bayes) risk in that one dimension. This cleanly separates the two questions — SSMD answers "which direction?" and the Bayes rule answers "where to cut?"

Project every subject onto the SSMD score $L = \boldsymbol{X}^T\,\widehat{\boldsymbol{\beta}}$. Within each group the score is (approximately) normal, with parameters estimated directly in score space:

$$L\big|G_1 \sim N\big(\widehat{\mu_{L1}}, \widehat{\sigma_{L1}^2}\big), \qquad L\big|G_2 \sim N\big(\widehat{\mu_{L2}}, \widehat{\sigma_{L2}^2}\big)$$

where $\widehat{\mu_{Lk}} = \widehat{\boldsymbol{\beta}}^{\boldsymbol{T}}\overline{\boldsymbol{X}}_{\boldsymbol{k}}$, $\widehat{\sigma_{Lk}^2} = \widehat{\boldsymbol{\beta}}^{\boldsymbol{T}}\mathbf{S}_{\boldsymbol{k}}\widehat{\boldsymbol{\beta}}$.

The Bayes classifier assigns the label with the larger posterior, i.e. the larger prior-weighted density, where the priors $\pi_1$, $\pi_2$ are the class proportions (such as 80/400 and 320/400):

$$\text{classify } L \text{ as } G_1 \quad <=> \quad \pi_1 f_1(L) > \pi_2 f_2(L)$$

$$\text{where } f_k = N\big(\widehat{\mu_{Lk}}, \widehat{\sigma_{Lk}^2}\big)$$

Taking logs and cancelling the common √(2π) factors, the decision boundary is the quadratic equation

$$\frac{(L - \widehat{\mu_{L2}})^2}{2\widehat{\sigma_{L2}^2}} - \frac{(L - \widehat{\mu_{L1}})^2}{2\widehat{\sigma_{L1}^2}} + \ln\frac{\pi_1}{\pi_2} - \frac{1}{2}\ln\frac{\widehat{\sigma_{L1}^2}}{\widehat{\sigma_{L2}^2}} = 0$$

This is the one-dimensional quadratic-discriminant rule written in score space. It has two roots; the cutoff *c* is the one lying between the group means. For the simulated data shown in Figure 2, the score-space statistics are $\widehat{\mu_{L1}}$ = 1.48, $\widehat{\mu_{L2}}$ = −1.56, $\widehat{\sigma_{L1}^2}$ = 0.43, $\widehat{\sigma_{L2}^2}$ = 4.11, $\pi_1$ = 0.20, $\pi_2$ = 0.80, which give *c* = +0.86 — matching the brute-force risk minimizer to four decimals.

**Reduction to the midpoint.** When the two score variances are equal ($\sigma_{L1}^2 = \sigma_{L2}^2 = \sigma_L^2$), the quadratic term vanishes and the boundary becomes linear:

$$c = \frac{\widehat{\mu_{L1}} - \widehat{\mu_{L2}}}{2} - \frac{\widehat{\sigma_L^2}}{\widehat{\mu_{L1}} - \widehat{\mu_{L2}}}\ln\frac{\pi_1}{\pi_2}$$

With equal priors, this collapses to the midpoint $\frac{\widehat{\mu_{L1}} - \widehat{\mu_{L2}}}{2}$ used by LDA as illustrated in Figure 1. So the midpoint is the Bayes cutoff under the (often violated) equal-variance, equal-prior assumption. As shown by Zhang [23], the middle point cutoff can also be derived by maximizing Youden index under the condition of normality with equal variance. The quadratic rule is what to use when those conditions fail, as in the example in Figure 2.

**Why integrate Bayes cutoff with SSMD.** Pairing the two combines complementary strengths and keeps a clean division of labor:

- **Separation of concerns.** SSMD finds the most-separated direction (an effect-size objective, robust and threshold-free), while the Bayes step sets the operating point. Neither has to compromise: the direction is chosen to maximize separation, the threshold to minimize error.
- **Handles heteroscedasticity and imbalance.** Because the Bayes cutoff uses the per-group variances and the priors, it automatically shifts toward the tighter, rarer group — correcting the LDA midpoint, which ignores both. In the example this lowers error from 18.5% to 13.0% with no change to the projection.
- **One-dimensional and stable.** After projection the cutoff is estimated from just four scalars ($\widehat{\mu_{L1}}$, $\widehat{\mu_{L2}}$, $\widehat{\sigma_{L1}^2}$, $\widehat{\sigma_{L2}^2}$) plus the priors, so it is far less data-hungry than fitting a full quadratic classifier in *k* dimensions, yet recovers most of the benefit when the SSMD direction is already near-optimal for separation.

- **Interpretable and tunable.** The cutoff has an explicit formula, and the priors can be replaced by costs: substituting $\pi_2/\pi_1$ with the false-positive/false-negative cost ratio yields the cost-optimal threshold, letting practitioners trade sensitivity against specificity by moving a single number.

### 5.2 A nonparametric classification cutoff based on the empirical ROC curve

The Bayes cutoff above is parametric: it assumes the projected score is Gaussian within each group and uses the estimated means, variances, and priors. A useful complementary choice drops the Gaussian assumption entirely and reads the cutoff straight off the empirical ROC curve. Sweep the threshold across the projected score and, at each value, record the true-positive rate (sensitivity) and false-positive rate; Youden's index J = sensitivity + specificity − 1 is maximized at the point on the ROC curve farthest from the chance diagonal. The score at that point is the Youden cutoff. Like the Bayes rule it is valid under heteroscedasticity (it never assumes equal variances, so it does not collapse to the midpoint), but unlike the Bayes rule it is nonparametric and prior-free: it weights the two error types equally and ignores the class proportions. That makes it a robust default when the within-group normality of the score is in doubt or when the operational costs are deliberately symmetric. Consequently, the Youden cutoff typically lies between the equal-variance midpoint cutoff and the Bayes-optimal cutoff. As illustrated later in Figure 3 and Table 3, it substantially improves upon the midpoint rule under heteroscedasticity while remaining distribution-free, although it generally does not achieve the minimum overall error attainable by the prior-aware Bayes decision rule.

## 6. Comparison with Fisher LDA, Logistic Regression and Linear SVM

### 6.1 Two Unique Features of SSMD: AUROC Relationship and Broad Applicability

Among the learning objectives discussed above, SSMD has two features that together distinguish it from Fisher LDA, logistic regression, and linear SVM. The first is its direct mathematical relationship with the area under the receiver operating characteristic curve (AUROC), one of the most widely used measures of classification performance. As summarized in Table 1 of Zhang [22], AUROC is mathematically identical to the d+-probability for all distributions, equals Φ(SSMD) under normality, and has explicit lower bounds as functions of SSMD for symmetric unimodal and general unimodal distributions with finite variance [7, 22, 24].

These relationships imply that maximizing SSMD is equivalent to maximizing AUROC under the Gaussian model and, more generally, increasing SSMD necessarily increases the lower bound of AUROC over a broad class of distributions. In contrast, Fisher LDA, logistic regression, and linear SVM optimize the Rayleigh quotient, conditional likelihood, and maximum margin, respectively, but none of these objectives has a direct analytical relationship with AUROC. This unique connection gives SSMD an important advantage as an interpretable learning objective: it simultaneously quantifies the separation between two classes and provides a probabilistic interpretation of classifier performance through AUROC.

**Table 1. Situations covered by SSMD maximization of a linear combination and the corresponding AUROC relationship for the resulting score**

| Variance / correlation structure | Distribution | SSMD-maximizing direction β* | AUROC relationship |
|---|---|---|---|
| *(i) Independent, equal covariance (situation a)* | Normal | $\boldsymbol{\Sigma}^{-1}(\boldsymbol{\mu}_1 - \boldsymbol{\mu}_2)$ coincides with Fisher LDA | AUROC = Φ(SSMD) |
| *(ii) Independent, unequal covariance (situation b)* | Normal | $(\boldsymbol{\Sigma}_1 + \boldsymbol{\Sigma}_2)^{-1}(\boldsymbol{\mu}_1 - \boldsymbol{\mu}_2)$ Anderson–Bahadur direction | AUROC = Φ(SSMD) |
| *(iii) Independent* | Symmetric unimodal, non-zero finite variance | $(\boldsymbol{\Sigma}_1 + \boldsymbol{\Sigma}_2)^{-1}(\boldsymbol{\mu}_1 - \boldsymbol{\mu}_2)$ | $\begin{cases} \text{AUROC} \geq 1 - \frac{2}{9\ \text{SSMD}^2}, & \text{when SSMD} \geq \sqrt{\frac{8}{3}} \\ \text{AUROC} \geq \frac{7}{6} - \frac{2}{3\ \text{SSMD}^2}, & \text{when } 1 \leq \text{SSMD} < \sqrt{\frac{8}{3}} \end{cases}$ |
| | Unimodal, non-zero finite variance | $(\boldsymbol{\Sigma}_1 + \boldsymbol{\Sigma}_2)^{-1}(\boldsymbol{\mu}_1 - \boldsymbol{\mu}_2)$ | $\begin{cases} \text{AUROC} \geq 1 - \frac{4}{9\ \text{SSMD}^2}, & \text{when SSMD} \geq \sqrt{\frac{8}{3}} \\ \text{AUROC} \geq \frac{4}{3} - \frac{4}{3\ \text{SSMD}^2}, & \text{when } 1 \leq \text{SSMD} < \sqrt{\frac{8}{3}} \end{cases}$ |
| *(iv) Correlated / paired (situation c)* | Normal | $\boldsymbol{\Sigma}_d^{-1}\boldsymbol{\mu}_d$ | AUROC = Φ(SSMD) with SSMD based on paired difference |
| *(v) Correlated / paired (situation c)* | Symmetric unimodal, non-zero finite variance | $\boldsymbol{\Sigma}_d^{-1}\boldsymbol{\mu}_d$ | $\begin{cases} \text{AUROC} \geq 1 - \frac{2}{9\ \text{SSMD}^2}, & \text{when SSMD} \geq \sqrt{\frac{8}{3}} \\ \text{AUROC} \geq \frac{7}{6} - \frac{2}{3\ \text{SSMD}^2}, & \text{when } 1 \leq \text{SSMD} < \sqrt{\frac{8}{3}} \end{cases}$ |
| | Unimodal, non-zero finite variance | $\boldsymbol{\Sigma}_d^{-1}\boldsymbol{\mu}_d$ | $\begin{cases} \text{AUROC} \geq 1 - \frac{4}{9\ \text{SSMD}^2}, & \text{when SSMD} \geq \sqrt{\frac{8}{3}} \\ \text{AUROC} \geq \frac{4}{3} - \frac{4}{3\ \text{SSMD}^2}, & \text{when } 1 \leq \text{SSMD} < \sqrt{\frac{8}{3}} \end{cases}$ |

**Key point:** In every row, maximizing the SSMD of the linear score maximizes AUROC under normality, or its guaranteed lower bound under unimodal distributions with finite variance. The SSMD-maximizing direction reduces to Fisher's LDA only in the independent, equal-covariance, normal case (row i); it becomes the Anderson–Bahadur direction under unequal covariances (rows ii–iii) and retains the cross-covariance $\Sigma_{12}$ in paired designs (rows iv–v), the only setting in which the pairing is exploited. Φ denotes the standard normal CDF; the unimodal lower bounds (symmetric-unimodal and general-unimodal) are the SSMD-increasing bounds of Zhang [7, 22, 24]. The table applies to positive SSMD values, and analogous results hold for negative SSMD values.

The second feature is the breadth of settings this objective covers, each with an AUROC reading. The three population forms of the SSMD in Section 2, together with the discriminant directions derived in Sections 6.2–6.4 below, let a single maximization principle span independent equal-covariance, independent heteroscedastic, and correlated (paired) designs under both normal and non-normal unimodal distributions. Table 1 collects these cases. For each combination of variance/correlation structure and distributional shape it gives the SSMD-maximizing direction $\boldsymbol{\beta}^*$ and the AUROC relationship that follows: the exact identity AUROC = $\Phi$(SSMD) under normality, and—for unimodal distributions with finite variance—explicit lower bounds on AUROC that increase with SSMD, in a tighter symmetric-unimodal form and a general-unimodal form.

## 6.2 Discriminant direction and cutoff under zero correlation and equal variance (situation a)

Fisher's linear discriminant analysis (LDA) seeks the projection that maximizes the ratio of between-class separation to within-class variation,

$$J(\boldsymbol{w}) = \frac{\boldsymbol{w}^T \boldsymbol{S}_{\mathrm{B}} \boldsymbol{w}}{\boldsymbol{w}^T \boldsymbol{S}_{\mathrm{W}} \boldsymbol{w}},$$

which yields the well-known solution $\boldsymbol{w}^* \propto \boldsymbol{S}_p^{-1}(\overline{\boldsymbol{X}}_1 - \overline{\boldsymbol{X}}_2)$, where $\boldsymbol{S_p} = \boldsymbol{S_W}$ is the pooled within-class covariance matrix. This direction is identical to the SSMD discriminant direction when the two classes share a common covariance matrix. Although SSMD and Fisher LDA are derived from different optimization criteria, they therefore produce the same discriminant direction under the equal-covariance assumption.

Table 2 summarizes the relationships among SSMD, Fisher LDA, logistic regression, and the linear support vector machine (SVM). Both SSMD and Fisher LDA determine the discriminant direction by multiplying the inverse pooled covariance matrix by the difference between the class mean vectors. Under the Gaussian equal-covariance model with equal class priors, they also use the same decision threshold—the midpoint between the projected class means—and consequently produce identical linear classifiers.

Logistic regression adopts a fundamentally different learning objective by maximizing the conditional likelihood of the class labels. Its decision boundary is defined by $\beta_0 + \boldsymbol{\beta}^{\mathrm{T}}\boldsymbol{X} = 0$, which corresponds to a predicted class probability of 0.5. Under the Gaussian equal-covariance model with equal class priors, this boundary coincides with the midpoint used by SSMD and Fisher LDA. However, when the class distributions overlap substantially or the class priors are unequal, the estimated intercept and regression coefficients adjust to minimize classification error, causing the decision boundary to shift away from the midpoint.

The linear SVM is based on yet another learning objective—the maximum-margin principle. Instead of using class means, covariance matrices, or likelihood functions, it determines the separating hyperplane $\boldsymbol{w}^{\mathrm{T}}\boldsymbol{X} + \boldsymbol{b} = 0$ by maximizing the margin between the two classes. Consequently, its solution is determined primarily by the support vectors closest to the decision boundary. Only under the symmetric Gaussian equal-covariance setting does the SVM hyperplane coincide with the midpoint classifier. In more general situations involving class overlap, unequal covariance structures, or class imbalance, the SVM boundary generally differs from those of SSMD, Fisher LDA, and logistic regression.

Overall, these comparisons highlight that SSMD is most closely related to Fisher LDA, sharing both the discriminant direction and, under standard Gaussian assumptions, the decision boundary. In contrast, logistic regression and linear SVM represent distinct machine-learning paradigms based on conditional probability estimation and margin maximization, respectively. Thus, while all four methods may yield equivalent classifiers under ideal distributional assumptions, they increasingly diverge as those assumptions become less restrictive.

**Table 2. Comparison of SSMD, Fisher Linear Discriminant Analysis, Logistic Regression, and Linear SVM: Discriminant Directions and Classification Boundaries**

| Method | Score direction β | Cutoff c (equal var, equal priors) |
|---|---|---|
| SSMD | $\boldsymbol{\Sigma}_*^{-1}(\boldsymbol{\mu}_1 - \boldsymbol{\mu}_2)$ | $(\mu_{L1} + \mu_{L2})/2$ (midpoint) |
| Fisher LDA | $S_p^{-1}(\boldsymbol{\mu}_1 - \boldsymbol{\mu}_2)$ | $(\mu_{L1} + \mu_{L2})/2$ (same midpoint) |
| Logistic | MLE of $\boldsymbol{\beta}$ (label likelihood) | where $\boldsymbol{\beta_0} + \boldsymbol{\beta}^{\mathbf{T}}\boldsymbol{X} = \mathbf{0}$, i.e. p = 0.5 |
| Linear SVM | w (max-margin normal) | where $\boldsymbol{w}^{\mathbf{T}}\boldsymbol{X} + \boldsymbol{b} = \mathbf{0}$ |

**Key point:** SSMD and Fisher LDA share the same discriminant direction and, under the Gaussian equal-covariance model with equal class priors, the same midpoint decision boundary, producing identical linear classifiers. Logistic regression yields the same boundary only under these assumptions but otherwise adjusts the boundary through likelihood maximization to account for class overlap and unequal priors. Linear SVM optimizes an entirely different objective by maximizing the geometric margin, making its decision boundary dependent primarily on the support vectors rather than on class means and covariance matrices. Consequently, SVM generally departs from the midpoint classifier except in symmetric, well-separated settings.

Below is a simulation to illustrate the score directions (i.e., projection in geometry) and classification boundaries summarized in Table 2. Two bivariate normal groups (n = 100 each, $\mu_1 = (1,1)$, $\mu_2 = (-1,-1)$, common covariance with correlation 0.5) illustrate the four methods. Panel (a) shows the projection directions; panels (b)–(e) show the projected-score histograms with the midpoint cutoff (dashed) and the resulting confusion matrix.

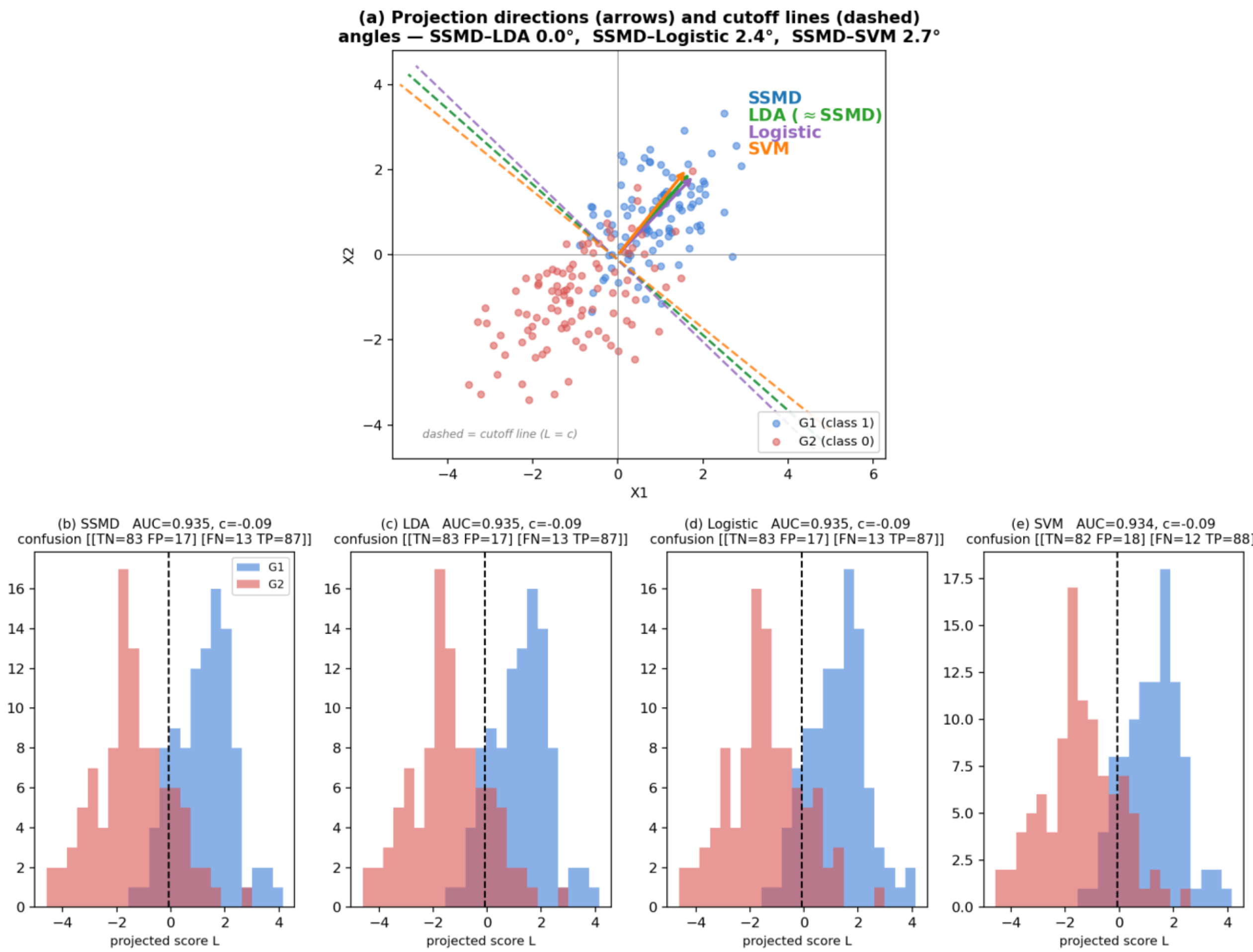

**Figure 1.** *Projection directions (arrows) and their corresponding cutoff lines (dashed, perpendicular to each arrow) for SSMD, Fisher LDA, logistic regression, and linear SVM on the same simulated data, with the projected-score histograms in panels (b)–(e). With independent groups and equal covariance the SSMD and LDA directions are numerically identical (0° apart), the logistic direction differs by 2.4°, and the SVM direction by 2.7°, so the four cutoff lines essentially coincide and pass through the overlap zone between the clouds; all four achieve AUC ≈ 0.93–0.94. The dashed line in (b)–(e) is the equal-variance midpoint cutoff of Section 5.1, which the 2D cutoff line in (a) renders geometrically.*

Because the data satisfy the equal-covariance, independent-group assumptions, SSMD and LDA coincide exactly, confirming Section 3.2. Logistic regression and the linear SVM nearly coincide here but would separate visibly under unequal covariances or class imbalance — as Figure 2 shows below. The confusion matrices at the shared midpoint cutoff are essentially identical across all four methods, matching the cutoff analysis of Section 5.1.

In summary

- SSMD — maximizes effect size; direction $\boldsymbol{\Sigma}_*^{-1}\,(\boldsymbol{\mu}_1 - \boldsymbol{\mu}_2)$; best for ranking and AUROC guided classification.
- LDA — equals SSMD under equal covariance and $\Sigma_{12}$ = 0; adds a classification rule and shares the midpoint cutoff.
- Logistic regression — maximizes label likelihood; robust under overlap/imbalance; cutoff at p = ½; yields calibrated probabilities.

- Linear SVM — maximizes the soft margin using only the points nearest the boundary; no calibrated effect size or confidence interval, but a useful assumption-light cross-check on the direction.

### 6.3 Discriminant direction and cutoff under zero correlation and unequal variance (situation b)

The example in Section 6.2 was deliberately benign — independent groups, equal covariance, balanced classes — so SSMD, LDA, logistic regression, and the linear SVM coincided in both direction and cutoff. That agreement is not generic. It breaks as soon as the assumptions that make them equivalent are violated. This section constructs a case that separates all four, and shows that the SVM's margin-based direction can land very close to SSMD's even when LDA's pooled-covariance direction does not.

Two ingredients drive the methods apart:

- **Unequal group covariances.** SSMD uses the sum of the two covariance matrices, whereas LDA uses the size-weighted pooled within-class scatter:

$$\Sigma_* = S_1 + S_2 \quad \text{vs.} \quad S_w = \frac{(n_1 - 1)S_1 + (n_2 - 1)S_2}{n_1 + n_2 - 2}$$

When $\Sigma_1 \neq \Sigma_2$ and the groups are imbalanced, these two matrices are no longer proportional, so $\mathbf{\Sigma}_*^{-1}\,(\boldsymbol{\mu}_1 - \boldsymbol{\mu}_2)$ and $\mathbf{S}_{\mathrm{w}}^{-1}\,(\boldsymbol{\mu}_1 - \boldsymbol{\mu}_2)$ point in different directions. Logistic regression, fitting the label likelihood, tilts differently again, and the linear SVM's margin-based direction need not match any of them — though Figure 2 shows it can land surprisingly close to SSMD's.

- **Unequal score variances.** Once the projected score has a different spread in each group, the optimal cutoff is no longer the midpoint of the means — it moves toward the tighter-variance group, by the quadratic rule of Section 5.1.

Concretely, take $G_1$ (class 1, $n_1$ = 80) with mean (1.5, 0) and covariance diag(0.5, 3), and $G_2$ (class 0, $n_2$ = 320) with mean (−1.5, 0) and covariance diag(4, 0.5). The clouds are elongated in perpendicular directions and the classes are imbalanced 1:4.

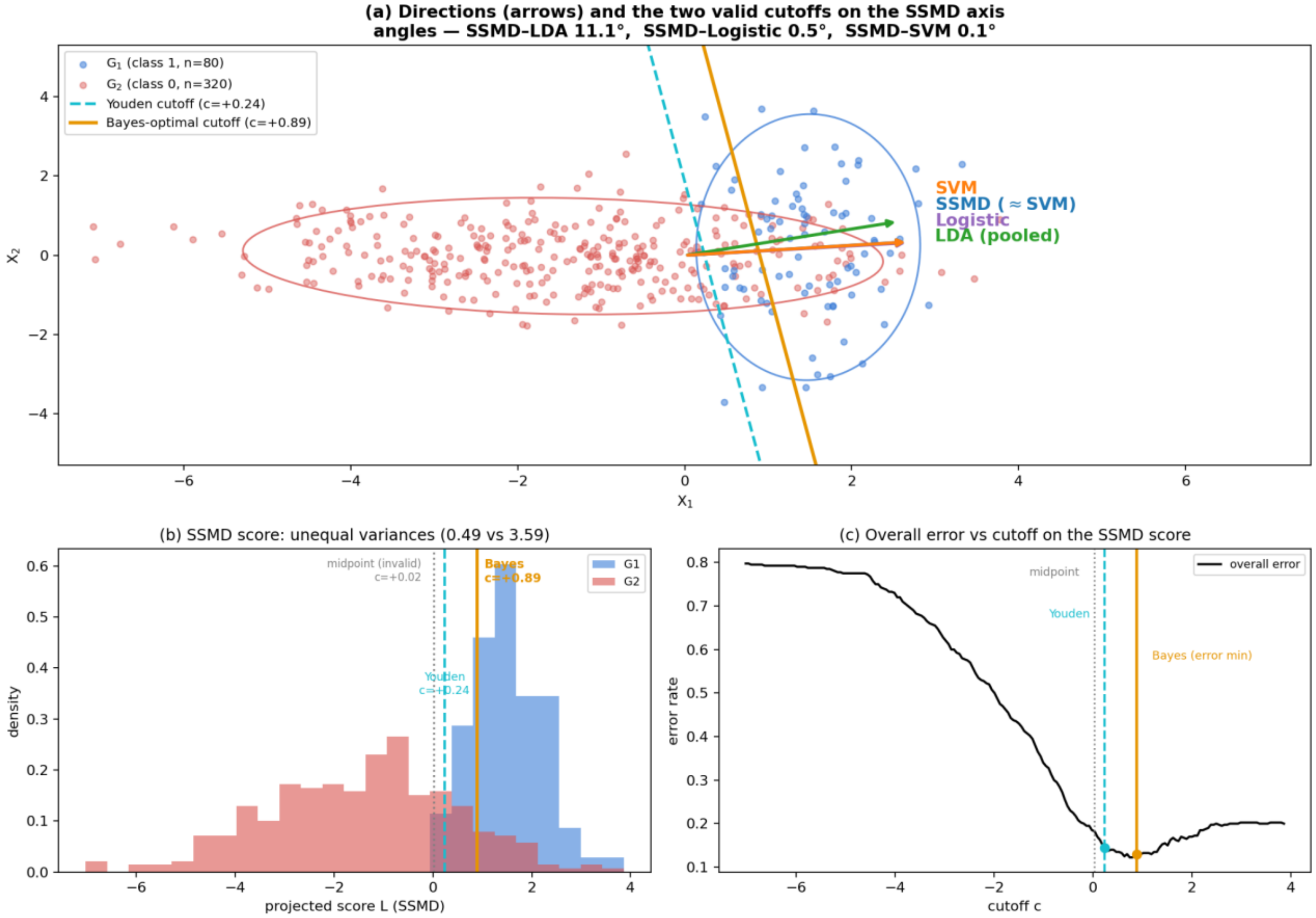

**Figure 2.** *A heteroscedastic, class-imbalanced example.* ***(a)*** *The four projection directions now genuinely differ in how far they tilt from SSMD (SSMD–LDA 11.1°, SSMD–Logistic 0.5°, SSMD–SVM 0.1°); the covariance ellipses show the conflicting cloud orientations that pull LDA's pooled direction away from the other three. Because the score variances are unequal, the equal-variance midpoint is no longer a valid operating point, so it is not drawn as a cutoff line; instead two heteroscedasticity-aware cutoffs are shown on the SSMD axis. The dashed cyan line is the Youden (empirical-ROC) cutoff and the solid orange line is the Bayes-optimal cutoff — both visibly shifted into the* $G_1$ *region, which is why they trim the wide, populous* $G_2$ *cloud and lower error, with the Bayes line shifted furthest.* ***(b)*** *On the SSMD score the two groups have very different variances (0.49 vs 3.59), so the equal-variance midpoint (+0.02, dotted, shown only for reference) is invalid; the Youden cutoff (+0.24, dashed cyan) and the Bayes-optimal cutoff (+0.89, orange) both sit well to its right.* ***(c)*** *The overall-error-versus-cutoff curve confirms the midpoint is far from the error minimizer; the Youden cutoff lands on the low-error shoulder and the Bayes-optimal cutoff sits at the minimum.*

The numbers below quantify the gap. The directions still produce nearly identical ranking quality (all AUC ≈ 0.93), because AUC is threshold-free — but the choice of cutoff materially changes the confusion matrix.

**Why the SVM direction nearly coincides with SSMD here.** The linear SVM [17] is a margin-based classifier: it chooses the direction and offset that maximize the soft margin, so its solution is governed by the support vectors near the boundary rather than by the full second-moment structure of each class. Because SSMD's objective is a ratio of mean separation to pooled dispersion and the SVM's is a hinge-loss margin, there is no algebraic reason for the two directions to coincide — yet here they land within 0.1° of each other, both well away from LDA's pooled-covariance direction. The reason is instructive: under unequal within-group variances, the wide-variance class spreads its points across the boundary region, so the support vectors anchoring the SVM margin are dominated by that class; tilting the direction to respect the unequal spread improves the margin in roughly the same way it improves the SSMD ratio. Logistic regression, which also adapts to the heteroscedastic likelihood, tracks the same direction. Fisher's LDA, by contrast, averages the two covariances into a single pooled scatter and so cannot represent the differing spreads, leaving it rotated off the SSMD optimum. Two qualifications keep this from being over-read. First, the near-coincidence is a property of this Gaussian, independent-group regime; it is not an identity — with strong outliers or heavy tails the SVM's support-vector focus and SSMD's moment-based ratio can pull apart, because a few boundary points move the margin without much changing the class means and variances. Second, the SVM optimizes margin, not effect size: it returns a direction and a threshold but no calibrated SSMD or its confidence interval. SSMD remains the method of choice when the deliverable is an interpretable, unit-free effect size with the inference of Section 4; the SVM is a useful, assumption-light cross-check that the SSMD direction is also a good classification direction.

Table 3. Comparison of Classification Cutoffs under Independent Unequal-Covariance Populations: Midpoint (LDA), Youden (ROC), and Bayes-Optimal Decision Rules

| Quantity | Midpoint cutoff (LDA) | Youden cutoff (ROC) | Bayes-optimal cutoff |
|---|---|---|---|
| Cutoff c on SSMD score | +0.02 | +0.24 | +0.89 |
| Confusion [TN, FP, FN, TP] | 247, 73, 1, 79 | 264, 56, 2, 78 | 286, 34, 18, 62 |
| Overall error rate | 18.5% | 14.5% | 13.0% |

**Reading the table.** The midpoint cutoff almost never misses a minority-class subject (FN = 1) but pays for it with 73 false positives, because it ignores that $G_2$ is both larger and far more spread out. The Youden cutoff moves partway right, already trimming false positives to 56 and lowering error to 14.5% without using the priors. The Bayes-optimal cutoff shifts further right, trading a little minority recall for many fewer false positives, and cuts overall error from 18.5% to 13.0%. This is precisely the regime where blindly using the LDA midpoint is suboptimal and where logistic regression and the linear SVM — which both adapt to the heteroscedastic likelihood and land within 0.5° of the SSMD direction here — or an explicit Bayes/QDA cutoff should be preferred over the pooled-covariance LDA boundary.

## 6.4 Discriminant direction and cutoff under non-zero correlation (situation c)

Figures 1 and 2 both assumed independent groups, so the cross-covariance $\Sigma_{12}$ vanished and SSMD's denominator reduced to $\Sigma_1+\Sigma_2$. When the two measurements are *paired* — the same subject assayed under two conditions, a pre/post design, or matched case–control units — the scores are correlated, $\Sigma_{12} \neq 0$, and the SSMD denominator keeps the full term $\mathbf{\Sigma}_* = \Sigma_1+\Sigma_2-\Sigma_{12}-\Sigma_{12}^\top$. This is the one regime where SSMD is structurally different from the other methods: Fisher's LDA uses the pooled within-group scatter, logistic regression uses the label likelihood, and the linear SVM uses the margin between the pooled samples, and **none of the** **three sees the pairing**. Figure 3 illustrates the consequence on a simulated paired design with unequal within-condition covariances.

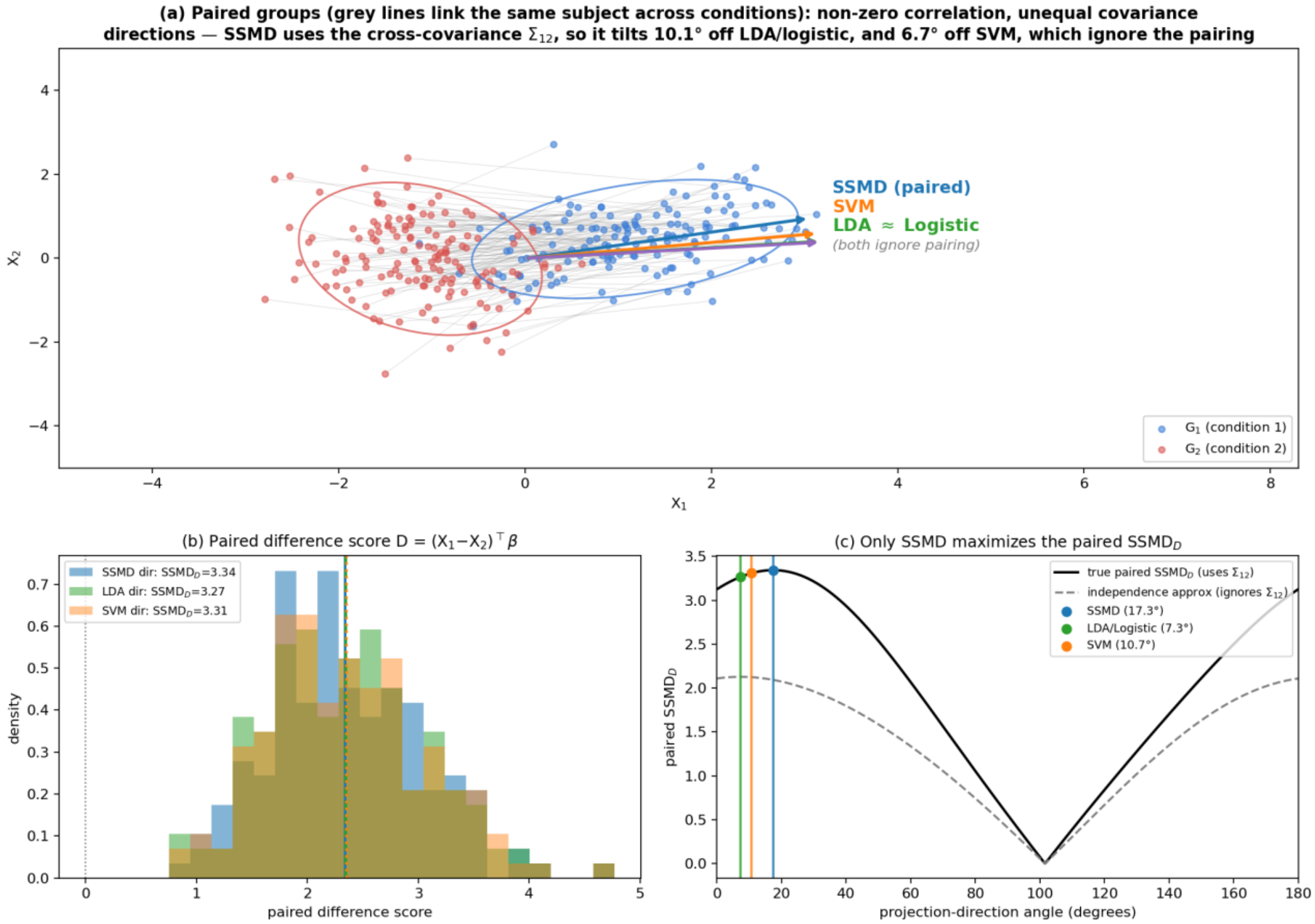

**Figure 3.** *A paired, correlated, heteroscedastic example (situation c).* ***(a)*** *The grey lines link each subject's two measurements, making the positive cross-covariance visible; the covariance ellipses differ in shape and orientation (unequal $\Sigma_1$, $\Sigma_2$). Because SSMD's denominator subtracts $\Sigma_{12}+\Sigma_{12}^T$, the SSMD direction tilts about 10.1° away from the LDA and logistic directions, which coincide here (0.3° apart) because both ignore the pairing; the linear SVM, which also ignores the pairing but is governed by the margin rather than the pooled covariance, lands 6.7° from SSMD — closer than LDA/logistic, but still off the paired optimum.* ***(b)*** *Distribution of the paired difference score $D = (X_1-X_2)^T\beta$: projecting on the SSMD direction yields the largest paired SSMD (3.34), against 3.31 for the SVM direction and 3.27 for the LDA/logistic direction.* ***(c)*** *The paired SSMD as a function of projection-direction angle peaks exactly at the SSMD solution; the SVM angle sits closer to the peak than the LDA/logistic angle, but both are off-peak. The grey dashed curve is the independence approximation that drops $\Sigma_{12}$ — it peaks at a different angle and underestimates the separation, quantifying what is lost by treating correlated groups as independent.*

The practical message is the converse of Figure 2. There, divergence was driven by unequal variance and class imbalance, and an explicit Bayes/QDA cutoff was the remedy. Here the direction itself diverges, and the gain is structural: only SSMD's objective contains $\Sigma_{12}$, so only SSMD recovers the direction that maximizes the paired effect size — LDA, logistic regression, and the linear SVM all ignore the pairing entirely, though the SVM's margin-based direction happens to land closer to SSMD's than the pooled-covariance methods do. As the correlation goes to zero this advantage disappears and situation (c) collapses back to situation (b); as the correlation grows, the gap between the SSMD direction and the pooled-scatter direction widens.

When a design is genuinely paired, estimating β with the cross-covariance term (and forming the confidence interval from the correlation-adjusted noncentral-*t* pivot of Section 4.2, situation c) is the statistically correct choice.

**In summary: when the four methods diverge**

- **Directions diverge** when the group covariances differ and the classes are imbalanced: SSMD (sum of covariances), LDA (pooled scatter), logistic regression (label likelihood), and the linear SVM (margin) each weight the data differently — though the SVM's margin-based direction can land surprisingly close to SSMD's under heteroscedasticity, as Figure 2 shows.
- **Cutoffs diverge** whenever the projected score has unequal variances across groups: the SSMD/LDA midpoint is no longer optimal, and the Bayes/QDA cutoff or logistic's p = ½ boundary can be substantially better.
- **Pairing is SSMD's exclusive advantage** (Figure 3): when the groups are correlated ($\Sigma_{12} \neq 0$), only SSMD's denominator keeps the cross-covariance, so only SSMD finds the direction that maximizes the paired effect size; LDA, logistic regression, and the linear SVM all ignore the pairing and tilt away from it, though the SVM tilts least.
- **The linear SVM is a useful but limited cross-check** (Figures 1–2): despite its unrelated margin-based objective, the SVM recovers a direction close to SSMD's in the Gaussian, independent-group regimes studied here (see Section 6.3 for why). This near-coincidence is a property of these regimes, not an algebraic identity, and the SVM returns a direction and threshold but no calibrated effect size or confidence interval—so SSMD remains preferable when the deliverable is an interpretable, unit-free effect size with the inference of Section 4.
- **Use SSMD** for direction-finding and effect-size ranking; use a Bayes/QDA cutoff or logistic regression when the goal is minimizing classification error under heteroscedasticity or imbalance.

# 7. Discussion

## 7.1 Summary of findings

This note has extended the strictly standardized mean difference from a single variable to a linear combination of variables, and has worked out the resulting estimation, inference, and classification theory. Three points stand out. First, maximizing the SSMD of a linear score is a generalized Rayleigh quotient whose solution has a closed form — the inverse of the covariance matrix of the difference between the two group scores applied to the vector of mean differences — and a clear relationship to classical discriminant analysis [16, 19]: it coincides with Fisher's direction when the two groups are independent with a common covariance, and departs from it otherwise. Second, the maximized SSMD can be estimated with the same small-sample care long applied to standardized effect sizes — method of moments, the bias-corrected uniformly minimum-variance unbiased estimator [7, 18], and maximum likelihood — and given exact or near-exact confidence intervals through the noncentral *t*-distribution [7]. Third, because the area under the ROC curve equals the standard normal cumulative probability evaluated at the SSMD [4, 22], the maximized score inherits an interpretable ROC summary and a principled operating point [21].

## 7.2 Principal merits of maximizing the SSMD of a linear combination

Beyond its closed-form solution and its small-sample inference, maximizing the SSMD of a linear combination carries four merits that, taken together, set it apart from the linear classifiers with which it is usually compared. All four flow from one fact: the SSMD of the linear score is analytically tied to the area under the ROC curve (AUROC).

*It maximizes AUROC, a threshold-free measure of overall discriminating capacity.* Under normality of the projected score, the SSMD and AUROC are linked by an exact identity: AUROC equals the standard normal cumulative probability evaluated at the SSMD [4, 22]. Because that cumulative probability increases strictly with its argument, the coefficient vector that maximizes the SSMD is exactly the one that maximizes AUROC. The distinction matters: AUROC integrates sensitivity and specificity over every possible cutoff, so it summarizes the overall discriminating capacity of the score rather than its accuracy at a single operating point. Maximizing SSMD therefore targets a cutoff-free measure of separation, and any linear classifier that recovers the SSMD-maximizing direction inherits the same reading — under the Gaussian model its projected score attains the largest AUROC available to any linear projection. Fisher's Rayleigh quotient, the conditional likelihood of logistic regression, and the SVM margin, by contrast, have no comparable closed-form relationship to AUROC.

*It gives Fisher's linear discriminant an AUROC interpretation.* When the two groups are independent with a common covariance, the SSMD-maximizing direction coincides with Fisher's discriminant (Sections 3.2 and 6.2), and the AUROC identity transfers with it. The discriminant that Fisher derived in 1936 as the linear function maximizing the ratio of between-group to within-group variation is, equivalently, the linear function that maximizes the area under the ROC curve under the equal-covariance Gaussian model. For many readers this characterization is more transparent and more useful than the classical variance ratio. A ratio of between-group to within-group variance is a mechanical quantity whose numerical value carries no direct probabilistic meaning; AUROC, on the zero-to-one scale familiar throughout the machine-learning and diagnostic literatures, is exactly the probability that the score ranks a randomly chosen subject from the first group above a randomly chosen subject from the second. The SSMD framework thus recasts Fisher's criterion in the language practitioners now use to report classifier performance.

*Its AUROC interpretation survives departures from normality.* The link to AUROC is not confined to the Gaussian case. For every distribution, AUROC equals the probability that a random draw from one group exceeds a random draw from the other [22]; and for symmetric unimodal distributions, and for general unimodal distributions with finite variance, this probability admits explicit lower bounds that are increasing functions of the SSMD [7, 22, 24]. Raising the SSMD of the linear score therefore raises a guaranteed floor on AUROC across a broad, non-normal class of distributions, and maximizing the SSMD maximizes that floor. This carries the discriminating-capacity interpretation well beyond the normal model — to the many biological readouts that are unimodal but skewed or heavy-tailed, where a normality-dependent criterion offers no such guarantee.

*It applies across a wide range of data-generating situations, each with an AUROC reading.* A single maximization principle spans independent equal-covariance, independent heteroscedastic, and correlated (paired) designs, under both normal and non-normal unimodal distributions; Table 1 (Section 6.1) sets out these cases with the discriminant direction and AUROC relationship each entails. Under normality the objective maximizes AUROC directly; under unimodal but possibly skewed or heavy-tailed distributions it maximizes a guaranteed lower bound on AUROC; and in paired designs its denominator alone retains the cross-covariance between matched measurements, so that among the four methods only SSMD maximization adapts its objective to the pairing. No competitor spans this range: Fisher's LDA is anchored to the equal-covariance case, and none of Fisher's LDA, logistic regression, or the linear SVM adapts to the paired cross-covariance or carries an AUROC guarantee into the non-normal unimodal regimes.

Together these properties make an interpretable, AUROC-anchored objective available with explicit small-sample inference across independent, heteroscedastic, paired, and non-normal settings alike — a combination that none of the compared linear classifiers provides.

## 7.3 Relationship to other linear methods

The simulations clarify when SSMD maximization agrees with, and when it departs from, the methods practitioners already use. Under independent groups with equal covariance, SSMD maximization, Fisher's LDA [16], and logistic regression [20] share the same population direction, AUROC, and Youden operating point [21]; divergence appears only under heteroscedasticity or class imbalance. In the heteroscedastic, imbalanced regime the optimal cutoff is no longer the midpoint of the score means and the methods' directions separate. In paired designs SSMD is distinctive: its denominator retains the cross-covariance between the two group scores, so it alone among the linear methods exploits the correlation between matched measurements. The margin-based linear support vector machine [17] behaves as an instructive modern comparator: although its hinge-loss objective is unrelated to an effect-size ratio, it empirically recovers nearly the SSMD direction in the Gaussian regimes studied here, while offering neither a calibrated effect size nor its confidence interval. SSMD maximization should therefore be seen not as a competitor that dominates these methods uniformly, but as the member of the family that targets an interpretable, unit-free effect size and supplies the matching inference [3, 5].

## 7.4 Limitations

Several limitations temper these results. First, the exact and near-exact confidence intervals of Section 4, and the identity linking AUROC to the SSMD, rest on approximate within-group normality of the projected score; when the score is markedly non-normal, the noncentral $F$ intervals lose their exactness, and the delta-method or bootstrap alternatives of Sections 4.3 and 4.4 are safer, as are the lower bounds established by Zhang [7, 22, 24] for the probability that a random draw from one group exceeds a random draw from the other — the quantity that equals AUROC — under unimodal distributions. Second, the optimal direction is a single linear projection: when the two groups have unequal covariances (situation b) the Bayes-optimal boundary is genuinely quadratic and no one direction is uniformly best, so the SSMD score is optimal only within the class of linear rules, and the Bayes/quadratic cutoff of Section 5.1 recovers part — but not all — of what a full quadratic classifier would (Section 6.3). Third, the optimal coefficients must be estimated, and plugging an estimated covariance inverse into the score can be unstable when the number of features is large relative to the sample size; that regime calls for regularized or sparse estimation of the coefficients, and, because reporting the SSMD on the same data used to fit them optimistically inflates it, the resampling interval of Section 4.4 is preferred. Finally, the comparisons here are drawn from Gaussian simulations designed to isolate specific effects; in particular, the near-coincidence of the SVM and SSMD directions should be read as regime-specific rather than as a general guarantee (Section 6.3).

A fifth limitation deserves separate emphasis, because it can bite even when the sample size is comfortably larger than the number of predictors. Both the optimal direction and the maximal SSMD are defined through the inverse of the covariance matrix of the difference between the two group scores, and are therefore sensitive to **collinearity** among the predictors — the rule rather than the exception in omics panels, high-content imaging features, and multiplexed biomarker assays, where many measurements report on the same underlying biology. When that matrix is ill-conditioned, its inverse amplifies sampling noise along the directions of least variation, with three consequences: the estimated direction becomes unstable, typically showing large, mutually cancelling weights on nearly redundant predictors that can no longer be read as measures of predictor importance; the plug-in maximal SSMD is inflated, the practical magnitude of the upward bias quantified in Section 4.1 being governed by the condition number rather than by the number of predictors alone; and in the limiting case of exact linear dependence the maximum is not attained at all, the sample analogue being the degeneracy that leaves Hotelling's statistic undefined once the degrees of freedom fall to or below the number of predictors. We therefore recommend that collinearity be assessed before a maximized-SSMD score is reported — the eigenvalue spectrum and condition number of that matrix, or variance-inflation factors, are adequate diagnostics — and that a regularized estimate be substituted when it is severe: ridge-type or convex shrinkage toward a well-conditioned target [20, 26], screening or sparse penalization of the predictor set [20], or projection onto leading principal components. Two caveats attach to any of these remedies. The regularized statistic no longer follows the Hotelling law, so the resampling interval of Section 4.4 — with the regularization refitted inside each resample — replaces the exact intervals of Section 4.2; and the amount of shrinkage should be chosen by cross-validation on a criterion external to the SSMD itself, since tuning it to maximize the observed SSMD reintroduces exactly the optimism that the interval is meant to quantify.

## 7.5 Applications and outlook

The framework is designed for settings where several measured features must be combined into one **interpretable**, well-separated score. These include hit selection and quality control in high-throughput screening [1, 2, 7, 9, 27–31], cytokine profiling [10], metabolomics, salivary biomarker panels for periodontitis and other diseases [11], and continuous-monitoring biomarkers for diabetes and respiratory research [12, 13]. In each, the maximized-SSMD signature provides a single score with a calibrated effect size, an AUROC interpretation [22], and a confidence interval, which together support thresholding, sample-size planning [15], and reporting. Several extensions are natural: regularized or sparse estimation of the coefficients when the number of features is large relative to the sample size [20], robust covariance estimation, and extension of the collective-activity idea [14] to signatures. I leave these to future work.

## ACKNOWLEDGMENTS

This work was supported by National Institutes of Health (AG084180, DK135111, GM156679), the University of Kentucky Barnstable Brown Diabetes and Obesity Center and the University of Kentucky Diabetes and Obesity Research Priority Area.

### Declaration of generative AI and AI-assisted technologies in the writing process

During the preparation of this work, the author used Claude in order to improve language and readability and coding. After using this tool, the author reviewed and edited the content as needed and take full responsibility for the content of the publication.

## References

**[1]** Zhang XD. A pair of new statistical parameters for quality control in RNA interference high-throughput screening assays. Genomics 2007; 89(4): 552–561.

**[2]** Zhang XD, Ferrer M, Espeseth AS, Marine SD, Stec EM, Crackower MA, Holder DJ, Heyse JF, Strulovici B. The use of strictly standardized mean difference for hit selection in primary RNA interference high-throughput screening experiments. J Biomol Screen 2007; 12(4): 497–509.

**[3]** Zhang XD. Strictly standardized mean difference, standardized mean difference and classical t-test for the comparison of two groups. Stat Biopharm Res 2010; 2(2): 292–299.

**[4]** Zhang XD. Illustration of SSMD, z-score, SSMD*, z*-score and t-statistic for hit selection in high-throughput screens. J Biomol Screen 2011; 16(7): 775–785.

**[5]** Zhang XD, Heyse JF. Contrast variable for comparing groups in biopharmaceutical research. Stat Biopharm Res 2012; 4(3): 228–239.

**[6]** Zhang XD. Contrast variable potentially providing a consistent interpretation to effect sizes. J Biom Biostat 2010; 1: 108.

**[7]** Zhang XD. Optimal High-Throughput Screening: Practical Experimental Design and Data Analysis for Genome-scale RNAi Research. Cambridge University Press, Cambridge, UK, 2011.

**[8]** Zhang XD, Lacson R, Yang R, Marine SD, McCampbell A, Toolan DM, Hare TR, Kajdas J, Berger JP, Holder DJ, Heyse JF, Ferrer M. The use of SSMD-based false discovery and false non-discovery rates in genome-scale RNAi screens. J Biomol Screen 2010; 15(9): 1123–1131.

**[9]** Zhang XD, Wang D, Sun S, Zhang H. Issues of z-factor and an approach to avoid them for quality control in high-throughput screening studies. Bioinformatics 2020; 36(22–23): 5299–5303.

**[10]** Saraswat S, Nurrahma BA, Kern PA, Nikolajczyk BS, Zhang XD. CytokineProfile: an integrated web tool dedicated to cytokine profiling analysis. Comput Struct Biotechnol J 2026; 35(1): Article 0079.

**[11]** Miller CS, Yan Q, Kirakodu SS, Ebersole JL, Zhang XD. Salivary biomarker panel that identifies periodontitis in persons with type 2 diabetes: a secondary analysis of a cross-sectional study. J Clin Periodontol 2026; 53(2): 190–200.

**[12]** Zhang T, Dong X, Chen C, Wang D, Zhang XD. RespirAnalyzer: an R package for continuous monitoring of respiratory signals. Bioinform Adv 2024; 4(1): vbae003.

**[13]** Zhang XD, Zhang Z, Wang D. CGManalyzer: an R package for analyzing continuous glucose monitoring studies. Bioinformatics 2018; 34(9): 1609–1611.

**[14]** Zhang XD, Santini F, Lacson R, Marine SD, Wu Q, Benetti L, Yang R, McCampbell A, Berger JP, Toolan DM, Stec EM, Holder DJ, Soper KA, Heyse JF, Ferrer M. cSSMD: assessing collective activity of multiple siRNAs in genome-scale RNAi screens. Bioinformatics 2011; 27(20): 2775–2781.

**[15]** Zhang XD, Heyse JF. Determination of sample size for hit selection in genome-scale RNAi screens. Bioinformatics 2009; 25(6): 841–844.

**[16]** Fisher RA. The use of multiple measurements in taxonomic problems. Ann Eugen 1936; 7(2): 179–188.

**[17]** Cortes C, Vapnik V. Support-vector networks. Mach Learn 1995; 20(3): 273–297.

**[18]** Hedges LV. Distribution theory for Glass's estimator of effect size and related estimators. J Educ Stat 1981; 6(2): 107–128.

**[19]** Anderson TW. An Introduction to Multivariate Statistical Analysis, 3rd ed. Wiley, Hoboken, NJ, 2003.

**[20]** Hastie T, Tibshirani R, Friedman J. The Elements of Statistical Learning: Data Mining, Inference, and Prediction, 2nd ed. Springer, New York, 2009.

**[21]** Youden WJ. Index for rating diagnostic tests. Cancer 1950; 3(1): 32–35.

**[22]** Zhang XD. Integrating AUROC and SSMD for quality control in high-throughput screening assays. SLAS Discov 2025; 36: 100269.

**[23]** Zhang XD. Hit Selection Using SSMD-Based Machine Learning Performance Metrics in High-Throughput Screening Assays. ArXiv 2026; 2608.07609

**[24]** Zhang XD. Novel Analytic Criteria and Effective Plate Designs for Quality Control in Genome-Scale RNAi Screens. J Biomol Screen 2008; 13(5): 363-377

**[25]** Nel DG, van der Merwe CA. A solution to the multivariate Behrens–Fisher problem. Commun Stat Theory Methods 1986; 15(12): 3719–3735.

**[26]** Ledoit O, Wolf M. A well-conditioned estimator for large-dimensional covariance matrices. J Multivar Anal 2004; 88(2): 365–411.

**[27]** Miedel MT, Varmazyad M, Xia M, Brooks MM, Gavlock DC, Reese C, Behari J, Soto-Gutierrez A, Gough A, Taylor DL, Schurdak ME. Validation of microphysiological systems for interpreting patient heterogeneity requires robust reproducibility analytics and experimental metadata. Cell Rep Methods 2025; 5(4): 101028.

**[28]** Meyers RM, Bryan JG, McFarland JM, et al. Computational correction of copy number effect improves specificity of CRISPR–Cas9 essentiality screens in cancer cells. Nat Genet 2017; 49(12): 1779–1784.

**[29]** Rossiter NJ, Huggler KS, Adelmann CH, et al. CRISPR screens in physiologic medium reveal conditionally essential genes in human cells. Cell Metab 2021; 33(6): 1248–1263.e9.

**[30]** Wheway G, Schmidts M, Mans DA, et al. An siRNA-based functional genomics screen for the identification of regulators of ciliogenesis and ciliopathy genes. Nat Cell Biol 2015; 17(8): 1074–1087.

**[31]** Yin J-A, Frick L, Scheidmann MC, et al. Arrayed CRISPR libraries for the genome-wide activation, deletion and silencing of human protein-coding genes. Nat Biomed Eng 2025; 9(1): 127–148.

**[32]** Nwizu C, Hughes M, Ramseier ML, Navia AW, Shalek AK, Fusi N, Raghavan S, Winter PS, Amini AP, Crawford L. Scalable nonparametric clustering with unified marker gene selection for single-cell RNA-seq data. Cell Rep Methods 2026; 6.